# Supersymmetric $E_{8(+8)}/SO(16)$ Sigma-Model Coupled to $N = 1$ Supergravity in Three-Dimensions

Hitoshi NISHINO[1] and Subhash RAJPOOT[2]

*Department of Physics & Astronomy*
*California State University*
*Long Beach, CA 90840*

## Abstract

A three-dimensional simple $N = 1$ supergravity theory with a supersymmetric sigma-model on the coset $E_{8(+8)}/SO(16)$ is constructed. Both bosons and fermions in the matter multiplets are in the spinorial **128**-representation of $SO(16)$ with the same chirality. Due to their common chirality, this model can not be obtained from the maximal $N = 16$ supergravity. By introducing an independent vector multiplet, we can also gauge an arbitrary subgroup of $SO(16)$ together with a Chern-Simons term. Similar $N = 1$ supersymmetric $\sigma$-models coupled to supergravity are also constructed for the cosets $F_{4(-20)}/SO(9)$ and $SO(8,n)/SO(8) \times SO(n)$.

[1] E-Mail: hnishino@csulb.edu
[2] E-Mail: rajpoot@csulb.edu

## 1. Introduction

There have been recently considerable developments in three-dimensional (3D) extended supergravity theories $2 \leq N \leq 16$ [1][2]. The $N = 16$ maximal supergravity theory has a non-trivial $\sigma$-model on the coset $E_{8(+8)}/SO(16)$ originally explored in [1], and its most general gaugings have been intensively studied in [2]. A more unified treatment of general extended supergravities in 3D has been also given in [3].

It has been well understood that other non-maximally extended supergravities with lower $N$ can be rather easily obtained by suitable truncations of the $N = 16$ maximal supergravity [1][3][2]. A typical example is $N = 12$ supergravity with the coset $E_{7(-5)}/SO(12) \times Sp(1)$ [1][3], which is a quaternionic Kähler manifold [4].

The maximal $N = 16$ supergravity [1][3][2] corresponds to 11D supergravity, and therefore it is supposed to be large enough to accommodate most of the possible target spaces coupled to lower $N$-extended as well as simple supergravity in 3D. Moreover, the spinorial **128**-representation of $SO(16)$ in $E_{8(+8)}/SO(16)$ is so peculiar that it seems difficult to couple such a coset to any lower extended supergravity theories. On the other hand, studies in 4D reveal that any Kähler manifold $\sigma$-model as the consistent target space coupled to $N = 1$ supergravity is further restricted to quaternionic Kähler manifold, when coupled to $N = 2$ supergravity [4]. Therefore, in 3D we expect an analogous restriction on $\sigma$-model target spaces. As a matter of fact, any Riemannian manifold can be the consistent target space coupled to $N = 1$ supergravity, while any Kähler manifold can be consistent with $N = 2$ supergravity [3]. The general possible target spaces for $\sigma$-models coupled to extended supergravities are categorized in [3] in an exhaustive and unified fashion, starting with $N = 1$ simple supergravity.

However, the formulation of $N = 1$ supersymmetric $\sigma$-models in [3] treats the matter fermions $\chi$ as the world-vector representation of the coset $G/H$, like the $\sigma$-model coordinate scalars. There is a subtlety about this treatment, because there is a difference between the chiral **128** and anti-chiral $\overline{\mathbf{128}}$-representation of $SO(16)$ for the fermions $\chi$, where the former was not covered as a special case in [3]. The method in [3] applied to the $N = 1$ as a special case is equivalent to the truncation of the $N = 16$ supergravity, in which the fermions will always come out to be in the anti-chiral $\overline{\mathbf{128}}$-representation.

There is another important motivation of studying the coset $E_{8(+8)}/SO(16)$ with $N = 1$ supersymmetry. In 3D, it seems true that $N = 1$ supersymmetric $\sigma$-model can be formulated without supergravity. If both local $N = 16$ supersymmetry and global global $N = 1$ supersymmetry can realize the supersymmetrization of the same coset $E_{8(+8)}/SO(16)$ with exactly the same physical degrees of freedom, then this provides a good motivation of studying such a globally $N = 1$ supersymmetric models. This link between global and local supersymmetries is associated with one important aspect of M-theory [5], namely, globally supersymmetric 1D matrix theory formulation [6][7] is supposed to reproduce 11D supergravity [8] with local supersymmetry, and this aspect may well be associated with the maximal coset $E_{8(+8)}/SO(16)$ in 3D. Additionally, a similar relation between local and global supersymmetries is found in the AdS/CFT correspondence, *i.e.*, the relationship between Type IIB superstring in 10D compactified into 5D and a globally $N = 4$ supersymmetric Yang-Mills theory in 4D [9][10], or analogous relationship between AdS(3) and CFT in 2D [11].

Motivated by these viewpoints, we will in this paper perform the coupling of supersymmetric $\sigma$-model on the coset $E_{8(+8)}/SO(16)$ to $N = 1$ supergravity, with the matter fermion in the spinorial **128**-representation. As has been mentioned, our methodology different from [3] is that both the $\sigma$-model scalars $\varphi_A$ and the fermions $\chi_A$ are in the **128**-representation of $SO(16)$.

We also conjecture that we can apply the same technique to other series of cosets, such as all the cosets given in Table 1 below.

| Supersymmetries | Cosets $G/H$ | $\varphi$ | $\chi$ |
|---|---|---|---|
| $N = 16$ | $E_{8(+8)}/SO(16)$ | **128** | **$\overline{128}$** |
| $N = 12$ | $E_{7(-5)}/SO(12) \times Sp(1)$ | **64** | **$\overline{64}$** |
| $N = 10$ | $E_{6(-14)}/SO(10) \times U(1)$ | **32** | **$\overline{32}$** |
| $N = 9$ | $F_{4(-20)}/SO(9)$ | **16** | **16** |

Table 1: Cosets and Representations Coupled to Extended Supergravities in 3D

The common feature of these cosets is that both the bosons $\varphi$ and fermions $\chi$ belong to the *spinorial* representation of $SO(n)$ in $H$. This conjecture is based on two main reasons: First, the coset $E_{8(+8)}/SO(16)$ is large enough to generate all of the cosets in Table 1 by its 'truncations'. Second, we can utilize the peculiar feature in 3D that both bosons and fermions can be in the same representations. In fact, we can overcome the difference in chirality, by choosing $\chi_A$ to be chiral (undotted) spinors instead of anti-chiral (dotted) spinors, as will be done for the cosets $E_{8(+8)}/SO(16)$ and $F_{4(-20)}/SO(9)$. As a by-product, another $N = 1$ supersymmetric $\sigma$-model on $SO(8, n)/SO(8) \times SO(n)$, which has been known in $N = 8$ supergravity [1], is coupled to supergravity. We will also add kinetic terms and a Chern-Simons term for a vector multiplet for gauging as a possible generalization.

## 2. $N = 1$ Supergravity with $\sigma$-Model on $E_{8(+8)}/SO(16)$

The construction of $N = 1$ supergravity shares many aspects with the $N = 16$ case [1][2] that we can take advantage of. However, we have also to pay attention to the difference of our formulation from the $N = 16$ case [1][2]. The most important difference is that the representation of our fermions $\chi_A$ belong to the chiral **128** of $SO(16)$. In ref. [3], $N = 1$ supersymmetric $\sigma$-models are understood as a special case of general $N$-extended supergravities. However, in that formulation, the fermions $\chi_{\dot{A}}$ always belong to the antichiral $\overline{\bf 128}$-representation, after appropriate truncations from $N \geq 2$ into $N = 1$.

To see how a different assignment is possible, we look into the linearized transformation of the latter for the fields $(\varphi_A, \chi_{\dot{A}})$:[3]

$$\delta_Q \varphi_A = +(\Gamma^I)_{A\dot{B}}(\bar{\epsilon}^I \chi_{\dot{B}}) \ , \quad \delta_Q \chi_{\dot{A}} = +\tfrac{i}{2}(\Gamma^I)_{B\dot{A}}\gamma^\mu \epsilon^I \partial_\mu \varphi_B \ . \tag{2.1}$$

The $A, B, \cdots = 1, 2, \cdots, 128$ and $\dot{A}, \dot{B}, \cdots = \dot{1}, \dot{2}, \cdots, \overline{128}$ are respectively the chiral and anti-chiral spinorial representations of $SO(16)$. Due to the index $I = 1, 2, \cdots, 16$ for the vectorial representation $\mathbf{16}$ of $SO(16)$ on $\epsilon^I$, we need to supply $\Gamma$-matrices in the r.h.s., in order to match the indices on the l.h.s. This structure is no longer imperative, when we need only $N=1$ supersymmetry. The simplest assignment is $(\varphi_A, \chi_A)$, with bosons and fermions in the same $\mathbf{128}$'s. Now the linear transformation rule is much simplified as

$$\delta_Q \varphi_A = +(\bar{\epsilon} \chi_A) \ , \quad \delta_Q \chi_A = +\tfrac{i}{2}\gamma^\mu \epsilon \partial_\mu \varphi_A \ . \tag{2.2}$$

Note that this special trick is possible, thanks to the peculiar feature of 3D supersymmetry, where the same numbers of bosonic and fermionic degrees of freedom can occupy exactly the same representation [12]. Notice also that within $N=16$, we can not put both $\varphi$ and $\chi$ in the same representation, due to the presence of the $\Gamma$-matrix saturating the $I$-index as in (2.1). Therefore, our multiplet $(\varphi_A, \chi_A)$ does not come from the maximal $N=16$ supergravity [1][2].

In ref. [3], a unified treatment of all the extended supergravities in 3D is presented including $N=1$ with the $\sigma$-model fermions $\chi_{\dot{A}}$ for $N=1$ in [3] in the anti-chiral $\overline{\mathbf{128}}$ of $H$ in $G/H$. The difference here is that the particular first Gamma matrix $(\Gamma^1)_{A\dot{B}}$ among the $\Gamma^I$-matrices is used to compensate the dottedness of each sides in [3], while our (2.2) does not need it. This is the reason our formulation is not covered as a special case in the unified formulation in [3].

Our next step is to couple such a $\sigma$-model to an $N=1$ supergravity multiplet. This is a routine construction, with Noether and quartic terms as in [1][2]. Our field content is $(e_\mu{}^m, \psi_\mu, \varphi_A, \chi_A)$, where $A, B, \cdots = 1, 2, \cdots, 128$ are for the $\mathbf{128}$ of $SO(16)$. The Maurer-Cartan form for coset representative $\mathcal{V}$ for the coset $E_{8(+8)}/SO(16)$ is [1][2]

$$\mathcal{V}^{-1}\partial_\mu \mathcal{V} = +P_{\mu A} Y_A + \tfrac{1}{2}Q_\mu{}^{IJ} X^{IJ} \ , \tag{2.3}$$

where $X^{IJ}$ and $Y_A$ are respectively the generators in $E_8$ in the directions of $SO(16)$ and the remaining coset $E_{8(+8)}/SO(16)$.

Our lagrangian for $N=1$ supergravity with the coset $E_{8(+8)}/SO(16)$ with a cosmological constant is rather simple:

$$\begin{aligned}\mathcal{L}_{E8} = & -\tfrac{1}{4}eR(e,\omega) + \tfrac{1}{2}\epsilon^{\mu\nu\rho}(\bar{\psi}_\mu D_\nu(\omega)\psi_\rho) + \tfrac{1}{4}eg^{\mu\nu}P_{\mu A}P_{\nu A} - \tfrac{i}{2}e(\bar{\chi}_A \gamma^\mu D_\mu(\omega,Q)\chi_A) \\ & -\tfrac{1}{2}e(\bar{\psi}_\mu \gamma^\nu \gamma^\mu \chi_A)P_{\nu A} - \tfrac{1}{8}e(\bar{\psi}_\mu \gamma^\nu \gamma^\mu \psi_\nu)(\bar{\chi}\chi) + \tfrac{1}{8}e(\bar{\chi}\chi)^2 - \tfrac{1}{96}e(\bar{\chi}\Gamma^{IJ}\gamma_\mu \chi)^2 \\ & + 2m^2 e + \tfrac{1}{2}me(\bar{\psi}_\mu \gamma^{\mu\nu}\psi_\nu) - \tfrac{1}{2}me(\bar{\chi}\chi) \ , \end{aligned} \tag{2.4}$$

---

[3]We are using the same notations as in [2] *e.g.*, the metric $(\eta_{mn}) = \text{diag.}\,(+,-,-)$, except for minor differences, such as putting all the spinorial (or vectorial) representation indices as subscripts (or superscripts), *etc.*.

whose action $I_{E8} \equiv \int d^3x \mathcal{L}_{E8}$ is invariant under supersymmetry

$$\delta_Q e_\mu{}^m = +i(\bar{\epsilon}\gamma^m \psi_\mu) \ , \quad \delta_Q \psi_\mu = +D_\mu(\omega)\epsilon + im\gamma_\mu \epsilon \ ,$$
$$\delta_Q \chi_A = +\tfrac{i}{2}\gamma^\mu \epsilon \widehat{P}_{\mu A} - \tfrac{1}{4}(\Gamma^{IJ}\chi)_A \Sigma^{IJ} \quad \left(\Sigma \equiv \tanh \tfrac{L}{2} S\right) \ ,$$
$$\delta_Q \varphi_A = +\frac{L}{\sinh L}(\bar{\epsilon}\chi_A) \equiv \left(\frac{L}{\sinh L} S\right)_A \ . \tag{2.5}$$

As in [1], $L$ acts on a function $f$ of $\varphi_A$ like $Lf \equiv [\mathcal{V}, f] \equiv [\exp(\varphi_A Y_A), f]$. The $\omega_\mu{}^{rs}$ is treated as an independent spinor connection as in the first-order formalism [13], and $R(e,\omega) \equiv e_m{}^\mu e_n{}^\nu R_{\mu\nu}{}^{mn}(\omega)$. Relevantly, we have the covariant derivative

$$D_\mu(\omega, Q)\chi_A \equiv \partial_\mu \chi_A + \tfrac{1}{4}\omega_\mu{}^{rs}\gamma_{rs}\chi_A + \tfrac{1}{4}Q_\mu{}^{IJ}(\Gamma^{IJ}\chi)_A \ . \tag{2.6}$$

Other notations, such as $(\overline{\chi}\chi) \equiv (\overline{\chi}_A \chi_A)$, are the same as in [1]. The positive coefficient for the cosmological constant in (2.4) implies anti-de Sitter background for our 3D supergravity, as desired.

Note also that our $N=1$ supergravity here can *not* be obtained from the maximal $N=16$ [1][2] by truncations covered in [3]. This is because both bosons and fermions of the $\sigma$-model are in the **128** of $SO(16)$, while those in [1][3][2] are in the chiral **128** and anti-chiral $\overline{\mathbf{128}}$. To put it differently, even though $N=1$ is much 'smaller' than $N=16$, it still maintains the same degrees of freedom $128+128$ as the $N=16$ theory [1][2], while the chirality assignment differentiates our theory from any descendants from the $N=16$ maximal supergravity [2]. In other words, there are two versions of $N=1$ supersymmetric $E_{8(+8)}/SO(16)$ $\sigma$-models coupled to supergravity, one with $\chi_A$ in the **128** as above, and another with $\chi_{\dot{A}}$ as in [3]. We can apply a similar method to the coset $F_{4(-20)}/SO(9)$ which will be performed in a more unified fashion in the next section.

## 3. Possibility of Gaugings

As for the gauging of subgroups of $E_8$ similar to that in $N=16$ [2], there seems to be some obstruction, due to the simple nature of $N=1$ supergravity. To be more specific, following [2], we set up the possible form of an additional lagrangian in terms of a vector $B_\mu{}^m$, and three functions $A_1$, $A_{2A}$ and $A_{3A}$ of the scalars $\varphi_A$:

$$\begin{aligned}\mathcal{L}_{E8}\big|_{g,B} \equiv &-\tfrac{1}{4}\epsilon^{\mu\nu\rho} B_\mu{}^m (\partial_\nu B_{\rho m} + \tfrac{1}{3}g f_{mnp} B_\nu{}^n B_\rho{}^p) \\ &+ \tfrac{1}{2}ge A_1(\overline{\psi}_\mu \gamma^{\mu\nu} \psi_\nu) + ige A_{2A}(\overline{\chi}_A \gamma^\mu \psi_\mu) + \tfrac{1}{2}ge A_{3AB}(\overline{\chi}_A \chi_B) \\ &+ a_1 g^2 e (A_1)^2 + a_2 g^2 e (A_{2A})^2 \ , \end{aligned} \tag{3.1}$$

with the minimal coupling constant $g$, and the transformation rule modifications

$$\delta_Q B_\mu{}^{\mathcal{M}} = +i\mathcal{V}_A{}^{\mathcal{M}}(\bar{\epsilon}\gamma_\mu \chi_A) \ ,$$
$$\delta_Q \psi_\mu\big|_g = +igA_1 \gamma_\mu \epsilon \ , \quad \delta_Q \chi_A\big|_g = +g A_{2A} \epsilon \ . \tag{3.2}$$

Note that due to the neutral gravitino $\psi_\mu$ under $SO(16)$, there is no $\psi$-linear term in $\delta_Q B_\mu$, in contrast to [2]. Accordingly, the Maurer-Cartan form (2.3) is modified to $\mathcal{V}^{-1}\mathcal{D}_\mu\mathcal{V} \equiv \mathcal{V}^{-1}\partial_\mu\mathcal{V} + \frac{1}{2}gB_\mu{}^{IJ}X^{IJ}\mathcal{V} \equiv \mathcal{P}_{\mu A}Y_A + \frac{1}{2}\mathcal{Q}_\mu{}^{IJ}X^{IJ}$ [2].

The first obstruction shows up in the commutator on $B_\mu{}^\mathcal{M}$, as the $g$-dependent term like $[\delta_1,\delta_2]B_\mu{}^\mathcal{M}|_g = +2ig\mathcal{V}_A{}^\mathcal{M}(\overline{\epsilon}_2\gamma_\mu\epsilon_1)A_{2A}$. The problem is that this term does not seem to be absorbed into the leading term that gives the desirable translation term $\approx (\overline{\epsilon}_2\gamma^\rho\epsilon_1)G_{\rho\mu}{}^\mathcal{M}$, by the use of the duality field equation relating $P_\mu$ to the field strength $G_{\mu\nu}{}^\mathcal{M}$ of $B_\mu{}^\mathcal{M}$. This forces us to impose the condition $A_{2A} = 0$, which in turn leads to the constancy of both $A_1$ and $A_{3AB}$, seen as follows: As in [2], we see that certain conditions to be satisfied by the $A$'s, categorized as the vanishing of the following terms in $\delta_Q\mathcal{L}_{\text{total}}$: (i) The $g\psi P$ or $g\psi DA_1$-terms. (ii) The $g\chi P$-terms. (iii) The $g^2\psi$-terms. (iv) The $g^2\chi$-terms. (v) The $g\psi\chi^2$-terms. These respectively yield the conditions

$$\mathcal{D}_\mu A_1 = A_{2A}\mathcal{P}_{\mu A} \ , \tag{3.3a}$$

$$\mathcal{D}_\mu A_{2A} - \tfrac{1}{2}A_1\mathcal{P}_{\mu A} - \tfrac{1}{2}A_{3AB}\mathcal{P}_{\mu B} + \tfrac{1}{2}T_{A|B}\mathcal{P}_{\mu B} = 0 \ , \tag{3.3b}$$

$$(a_1 - 2)(A_1)^2 + (a_2 + 1)(A_{2A})^2 = 0 \ , \tag{3.3c}$$

$$(2a_1 + a_2 - 3)A_1 A_{2A} + (a_2 + 1)A_{2A}A_{3AB} + a_2 T_{A|B}A_{2B} = 0 \ , \tag{3.3d}$$

$$A_1\delta_{AB} + A_{3AB} = 0 \ , \tag{3.3e}$$

where $T_{\mathcal{A}|\mathcal{B}} \equiv \mathcal{V}_\mathcal{A}{}^m\Theta_{mn}\mathcal{V}_\mathcal{B}{}^n$ in the notation in [2]. Once $A_{2A} = 0$ is accepted, then from (3.3a) the only non-trivial solutions seem to be $A_1 = \text{const.}$, $A_{2A} = 0$, which with (3.3c) and (3.3e) imply that $a_1 = +2$ and $A_{3A} = -\delta_{AB}A_1$. At this stage we have

$$A_1 = \text{const.} \neq 0. \ , \quad A_{2A} = 0 \ , \quad A_{3AB} = -\delta_{AB}A_1 \ , \quad a_1 = +2 \ . \tag{3.4}$$

The only remaining condition is (3.3b) with the last term left over, which comes originally from the minimal coupling between $B_\mu{}^\mathcal{M}$ and $\varphi_A$-scalars, implying that $T_{A|B} = 0$. Therefore, if we do not require the minimal couplings, all of the conditions in (3.3) are satisfied. For an obvious reason, we no longer need $a_2$, and this is nothing other than our result (2.4) with the cosmological constant with $m \equiv gA_1$.

This result indicates that there is no allowed gauge group in this formulation in contrast to $N = 16$ [2]. In a certain sense, this is reasonable, because $N = 1$ supergravity does not have so much freedom as $N = 16$ supergravity, and this may be the price to be paid. However, as will be mentioned, our method for $E_{8(+8)}/SO(16)$ can be easily applied to other cosets, such as those in Table 1. In that sense, even this $N = 1$ simple supergravity still has rich enough structures, when coupled to non-trivial $\sigma$-models. To put it differently, in $N = 1$ supergravity, we have more freedom to choose the coset out of certain series such as those in Table 1 all of which can consistently couple to $N = 1$ supergravity. In this sense, we do not need the freedom of 'gauging' that plays an important role in the case of $N = 16$ maximal supergravity [2]. Since the supergravity multiplet does not have any physical degrees of freedom in 3D, the actual difference in the number of supersymmetries $N$ may not be so crucial in 3D as analogous supergravities in dimensions 4D or higher.

The gauging we have failed is an 'internal' gauging, where $B_\mu{}^{IJ}$ transformed into $\chi$, with no independent gaugino $\lambda$. We can show, however, that certain 'external' gauging with an additional multiplet $(A_\mu{}^{(r)}, \lambda^{(r)})$ like the $N=1$ $\sigma$-model in 2D [14] is indeed possible. For simplicity, we consider only the cosets $G/H = E_{8(+8)}/SO(16)$ or $F_{4(-20)}/SO(9)$ in a more unified notation until the end of this section. For such a purpose, it is more convenient to switch to a notation using the metric $g_{\alpha\beta}$ on $G/H$, instead of using coset representatives $\mathcal{V}$ as we did above. The metric $g_{\alpha\beta} \equiv V_{\alpha A} V_{\beta A}$ is defined by the vielbein $V_{\alpha A}$ in the Maurer-Cartan form for the gauged case:

$$\mathcal{V}^{-1}\mathcal{D}_\mu\mathcal{V} \equiv \mathcal{V}^{-1}\partial_\mu\mathcal{V} + gA_\mu{}^{(r)}\mathcal{V}^{-1}T^{(r)}\mathcal{V} = (\mathcal{D}_\mu\phi^\alpha)V_\alpha{}^A Y_A + \tfrac{1}{2}(\mathcal{D}_\mu\phi^\alpha)Q_\alpha{}^{IJ}X^{IJ} \ ,$$
$$\mathcal{D}_\mu\phi^\alpha \equiv \partial_\mu\phi^\alpha - gA_\mu{}^{(r)}\xi^{\alpha(r)} \ , \tag{3.5}$$

with the curved indices $\alpha, \beta, \cdots = 1, 2, \cdots, \dim(G/H)$, and spinorial indices $A, B, \cdots = 1, 2, \cdots, d_S$ where $d_S$ is the dimensionality of the spinorial representation of $H$. The indices $(r), (s), \cdots = (1), (2), \cdots, (d_0)$ on the generators $T^{(r)}$ are for the adjoint representation of an arbitrary gauged subgroup $H_0 \subset H$ with $d_0 \equiv \dim H_0$. Accordingly, we use $\phi^\alpha$ for the coordinates for the coset $E_{8(+8)}/SO(16)$ or $F_{4(-20)}/SO(9)$, which are the same as $\varphi_A$ at the linear order, but different at higher orders.

Our total field content is now $(e_\mu{}^m, \psi_\mu, \phi^\alpha, \chi_A, A_\mu{}^{(r)}, \lambda^{(r)})$ for $N=1$ supergravity coupled to $\sigma$-model on $E_{8(+8)}/SO(16)$ or $F_{4(-20)}/SO(9)$ with an 'external' gauging. As an additional generalization, we can also add a Chern-Simons term for the field strength $F_{\mu\nu}{}^{(r)}$ [15]. The corresponding lagrangian is

$$\begin{aligned}\mathcal{L}_{G/H} =\ & -\tfrac{1}{4}eR(e,\omega) + \tfrac{1}{2}\epsilon^{\mu\nu\rho}(\overline{\psi}_\mu D_\nu(\omega)\psi_\rho) \\ & + \tfrac{1}{4}eg_{\alpha\beta}g^{\mu\nu}(\partial_\mu\phi^\alpha)(\partial_\nu\phi^\beta) - \tfrac{i}{2}e(\overline{\chi}_A\gamma^\mu\mathcal{D}_\mu(\omega,Q,A)\chi_A) \\ & - \tfrac{1}{4}eg^{\mu\nu}g^{\rho\sigma}F_{\mu\rho}{}^{(r)}F_{\nu\sigma}{}^{(r)} - ie(\overline{\lambda}^{(r)}\gamma^\mu D_\mu(\omega,A)\lambda^{(r)}) \\ & - \tfrac{1}{2}e(\overline{\psi}_\mu\gamma^\nu\gamma^\mu\chi_A)V_{\alpha A}\mathcal{D}_\nu\phi^\alpha - \tfrac{i}{2}e(\overline{\psi}_\mu\gamma^{\rho\sigma}\gamma^\mu\lambda^{(r)})F_{\rho\sigma}{}^{(r)} - ge(\overline{\chi}_A\lambda^{(r)})V_{\alpha A}\xi^{\alpha(r)} \\ & - \tfrac{1}{8}e(\overline{\psi}_\mu\gamma^\nu\gamma^\mu\psi_\nu)(\overline{\chi}_A\chi_A) + \tfrac{1}{8}e(\overline{\chi}_A\chi_A)^2 + \tfrac{1}{12}eR_{ABCD}(\overline{\chi}_A\gamma_\mu\chi_B)(\overline{\chi}_C\gamma^\mu\chi_D) \\ & + \tfrac{1}{2}e(\overline{\psi}_\mu\psi^\mu)(\overline{\lambda}^{(r)}\lambda^{(r)}) + \tfrac{1}{2}e(\overline{\lambda}^{(r)}\lambda^{(r)})^2 + \tfrac{1}{2}e(\overline{\chi}_A\chi_A)(\overline{\lambda}^{(r)}\lambda^{(r)}) \\ & + \tfrac{1}{2}cm\,\epsilon^{\mu\nu\rho}(F_{\mu\nu}{}^{(r)}A_\rho{}^{(r)} - \tfrac{1}{3}gf^{(r)(s)(t)}A_\mu{}^{(r)}A_\nu{}^{(s)}A_\rho{}^{(t)}) \\ & + 2m^2 e + \tfrac{1}{2}me(\overline{\psi}_\mu\gamma^{\mu\nu}\psi_\nu) - \tfrac{1}{2}me(\overline{\chi}_A\chi_A) + (2c-1)me(\overline{\lambda}^{(r)}\lambda^{(r)}) \ , \end{aligned} \tag{3.6}$$

whose action is invariant under supersymmetry

$$\begin{aligned} \delta_Q e_\mu{}^m &= +i(\overline{\epsilon}\gamma^m\psi_\mu) \ , & \delta_Q\psi_\mu &= +D_\mu(\omega)\epsilon + im\gamma_\mu\epsilon \ , \\ \delta_Q\phi^\alpha &= +V_A{}^\alpha(\overline{\epsilon}\chi_A) \ , & \delta_Q\chi_A &= +\tfrac{i}{2}\gamma^\mu\epsilon V_{\alpha A}\widehat{\mathcal{D}}_\mu\phi^\alpha - \tfrac{1}{4}(\delta_Q\phi^\alpha)Q_\alpha{}^{IJ}(\Gamma^{IJ}\chi)_A \ , \\ \delta_Q A_\mu{}^{(r)} &= +i(\overline{\epsilon}\gamma_\mu\lambda^{(r)}) \ , & \delta_Q\lambda^{(r)} &= -\tfrac{1}{4}\gamma^{\mu\nu}\epsilon\widehat{F}_{\mu\nu}{}^{(r)} \ . \end{aligned} \tag{3.7}$$

The penultimate line of (3.6) has a coefficient with an arbitrary constant $c$ that might be determined by the quantization of the Chern-Simons term. This constant enters also the gaugino mass term. The covariant derivative $\mathcal{D}_\mu$ contains both the composite and minimal couplings:

$$\begin{aligned}\mathcal{D}_\mu(\omega,Q,A)\chi_A \equiv\ & +\partial_\mu\chi_A + \tfrac{1}{4}\omega_\mu{}^{rs}\gamma_{rs}\chi_A \\ & + \tfrac{1}{4}(\mathcal{D}_\mu\phi^\alpha)Q_\alpha{}^{IJ}(\Gamma^{IJ}\chi)_A + 2gA_\mu{}^{(r)}(D_A(Q)\xi_B{}^{(r)})\chi_B\ ,\end{aligned} \quad (3.8)$$

The covariant derivative $D_A(Q)$ acts on the $\xi$'s like $D_A(Q)\xi_B{}^{(r)} \equiv V_A{}^\alpha[\partial_\alpha\xi_B{}^{(r)} + (1/4)Q_\alpha{}^{IJ}(\Gamma^{IJ})_{AB}\xi_B{}^{(r)}]$. Relevantly, the curvature tensor $R_{ABCD}$ is defined by

$$R_{ABCD} \equiv V_A{}^\alpha V_B{}^\beta R_{\alpha\beta CD} \equiv \tfrac{1}{4}V_A{}^\alpha V_B{}^\beta R_{\alpha\beta}{}^{IJ}(\Gamma^{IJ})_{CD}\ , \quad (3.9)$$

with the curvature tensor $R_{\alpha\beta}{}^{IJ}$ for the composite connection $Q_\alpha{}^{IJ}$ for the isotropy group $H$ [3]. More explicitly, both for the cosets $E_{8(+8)}/SO(16)$ and $F_{4(-20)}/SO(9)$, we have [3] [4]

$$R_{ABCD} = -\tfrac{1}{8}(\Gamma^{IJ})_{AB}(\Gamma^{IJ})_{CD}\ , \quad (3.10)$$

independent of the value $n$ in $SO(n)$ in $H$. This notation thus unifies the previous cases of $E_{8(+8)}/SO(16)$ and $F_{4(-20)}/SO(9)$. A crucial relationship in the invariance confirmation of our action is

$$D_A(Q)\xi_B{}^{(r)} = +\tfrac{1}{4}(\Gamma^{IJ})_{AB}Q_\alpha{}^{IJ}\xi^{\alpha(r)}\ , \quad (3.11)$$

which is proven by other relationships, such as $\xi_A{}^{(r)} = +(1/60)\,\text{tr}\,(\mathcal{V}^{-1}T^{(r)}\mathcal{V}Y_A)$.

The $g$-linear $(\overline\chi\lambda)$-term in (3.6) is much like similar terms in 2D heterotic $\sigma$-model [14] or $N=2$ gauged hypermultiplet couplings in 6D [16]. Compared with [16], due to the neutral gravitino under $H_0$, we have no mixture term between $\psi_\mu$ and $\lambda$ linear in $g$. The Chern-Simons term with the kinetic term for the vector field leads to the $A_\mu$-field equation

$$D_\nu F^{\mu\nu\,(r)} \doteq c\,m\,e^{-1}\epsilon^{\mu\nu\rho}F_{\nu\rho}{}^{(r)} + J^{\mu\,(r)}\ , \quad (3.12)$$

which is called 'generalized self-duality' condition in odd dimensions [17].

Since we no longer have strong restriction by the 'internal' gauging, there is no condition on the allowed gauge group $H_0$ as a subgroup of $SO(16)$ for $E_{8(+8)}/SO(16)$ or of $SO(9)$ for $F_{4(-20)}/SO(9)$. Additionally, the $N=1$ case has more freedom than $N=16$, for choosing the gravitino mass $m$ independently of the minimal gauge coupling constant $g$. In this sense, this $N=1$ system is much closer to the $N=1$ case in 2D [14] than $N=16$ in 3D, except that the former forbids the cosmological constant due to chirality. We can apply similar methods to other cosets in Table 1, which are to be skipped in this paper.

---

[4]The no sign-flip in our definition (3.9) compared with [3] costed an extra minus sign in (3.10).

## 4. Application to Coset $SO(8,n)/SO(8) \times SO(n)$

We have so far realized a supersymmetric $\sigma$-model with the non-trivial representations $\mathbf{128 + 128}$ on such a huge coset $E_{8(+8)}/SO(16)$ with simple $N=1$ supergravity, which is different from the formulation in [3]. A next natural question is whether this formulation is possible only with the cosets $E_{8(+8)}/SO(16)$ and $F_{4(-20)}/SO(9)$, or are there others? In fact, this question can be answered in the affirmative, shown by a similar construction for $SO(8,n)/SO(n) \times SO(8)$ which is not contained in Table 1, but has been known as a consistent coset for $N=8$ supergravity [1]. The common feature here is that the fermions on $SO(8,n)/SO(n) \times SO(8)$ are also in the spinorial $\mathbf{8}_S$ of $SO(8)$.

There is also a slight difference between $SO(8,n)/SO(n) \times SO(8)$ and $G/H = E_{8(+8)}/SO(16)$ or $F_{4(-20)}/SO(9)$, due to the additional isotropy group $SO(n)$ that needs additional care. We use a notation similar to [1], and assign the $(\mathbf{8}_S, \mathbf{n})$-representation of $SO(8) \times SO(n)$ to the fermions $\chi_{Aa}$. Accordingly, our field content is $(e_\mu{}^m, \psi_\mu, \phi^\alpha, \chi_{Aa}, A_\mu{}^{(r)}, \lambda^{(r)})$, where $A, B, \cdots = 1, 2, \cdots, 8$ are for the chiral $\mathbf{8}_S$ of $SO(8)$, and $a, b, \cdots = 1, 2, \cdots, n$ for the vectorial $\mathbf{n}$ of $SO(n)$. The difference here from refs. [1][3] is that our $\chi_{Aa}$ is in the $\mathbf{8}_S$ but not the $\mathbf{8}_C$-representation of $SO(8)$.

Our result for $N=1$ supergravity with the $SO(8,n)/SO(n) \times SO(8)$ is summarized by the lagrangian

$$\begin{aligned}
\mathcal{L}_{SO(8,n)} = &-\tfrac{1}{4} e R(e,\omega) + \tfrac{1}{2}\epsilon^{\mu\nu\rho}(\overline{\psi}_\mu D_\nu(\omega)\psi_\rho) \\
&+ \tfrac{1}{4} e g_{\alpha\beta} g^{\mu\nu}(\partial_\mu \phi^\alpha)(\partial_\nu \phi^\beta) - \tfrac{i}{2} e(\overline{\chi}_{Aa}\gamma^\mu \mathcal{D}_\mu(\omega,Q,A)\chi_{Aa}) \\
&- \tfrac{1}{4} e g^{\mu\nu} g^{\rho\sigma} F_{\mu\rho}{}^{(r)} F_{\nu\sigma}{}^{(r)} - i e(\overline{\lambda}^{(r)}\gamma^\mu D_\mu(\omega,A)\lambda^{(r)}) \\
&- \tfrac{1}{2} e(\overline{\psi}_\mu \gamma^\nu \gamma^\mu \chi_{Aa}) V_{\alpha Aa} \mathcal{D}_\nu \phi^\alpha - \tfrac{i}{2} e(\overline{\psi}_\mu \gamma^{\rho\sigma}\gamma^\mu \lambda^{(r)})F_{\rho\sigma}{}^{(r)} - ge(\overline{\chi}_{Aa}\lambda^{(r)})V_{\alpha Aa}\xi^{\alpha(r)} \\
&- \tfrac{1}{8} e(\overline{\psi}_\mu \gamma^\nu \gamma^\mu \psi_\nu)(\overline{\chi}_{Aa}\chi_{Aa}) + \tfrac{3}{32} e(\overline{\chi}_{Aa}\chi_{Aa})^2 - \tfrac{1}{64} e(\overline{\chi}\gamma_\mu \Gamma^{IJ}\chi)^2 - \tfrac{1}{1536} e(\overline{\chi}\Gamma^{IJKL}\chi)^2 \\
&+ \tfrac{1}{2} e(\overline{\psi}_\mu \psi^\mu)(\overline{\lambda}^{(r)}\lambda^{(r)}) + \tfrac{1}{2} e(\overline{\lambda}^{(r)}\lambda^{(r)})^2 + \tfrac{1}{2} e(\overline{\chi}_{Aa}\chi_{Aa})(\overline{\lambda}^{(r)}\lambda^{(r)}) \\
&+ \tfrac{1}{2} c m\, \epsilon^{\mu\nu\rho}(F_{\mu\nu}{}^{(r)}A_\rho{}^{(r)} - \tfrac{1}{3} g f^{(r)(s)(t)} A_\mu{}^{(r)} A_\nu{}^{(s)} A_\rho{}^{(t)}) \\
&+ 2m^2 e + \tfrac{1}{2} m\, e\, (\overline{\psi}_\mu \gamma^{\mu\nu}\psi_\nu) - \tfrac{1}{2} m e(\overline{\chi}_{Aa}\chi_{Aa}) + (2c-1)me(\overline{\lambda}^{(r)}\lambda^{(r)}) \quad , \quad (4.1)
\end{aligned}$$

and the supersymmetry transformation rule

$$\delta_Q e_\mu{}^m = +i(\overline{\epsilon}\gamma^m \psi_\mu) \;, \quad \delta_Q \psi_\mu = +D_\mu(\omega)\epsilon + im\gamma_\mu \epsilon \;,$$

$$\delta_Q \phi^\alpha = +V_{Aa}{}^\alpha (\overline{\epsilon}\chi_{Aa}) \;, \quad \delta_Q \chi_{Aa} = +\tfrac{i}{2}\gamma^\mu \epsilon V_{\alpha Aa}\widehat{\mathcal{D}}_\mu \phi^\alpha - (\delta_Q \phi^\alpha)\left[\tfrac{1}{4} Q_\alpha{}^{IJ}(\Gamma^{IJ}\chi_a)_A + Q_{\alpha ab}\chi_{Ab}\right] \;,$$

$$\delta_Q A_\mu{}^{(r)} = +i(\overline{\epsilon}\gamma_\mu \lambda^{(r)}) \;, \quad \delta_Q \lambda^{(r)} = -\tfrac{1}{4}\gamma^{\mu\nu}\epsilon \widehat{F}_{\mu\nu}{}^{(r)} \;. \quad (4.2)$$

Our notation is much like that in [1], and other relevant relationships are also similar to the previous section, such as the covariant derivative

$$\begin{aligned}
D_\mu(\omega,Q)\chi_{Aa} \equiv\; &+\partial_\mu \chi_{Aa} + \tfrac{1}{4}\omega_\mu{}^{rs}\gamma_{rs}\chi_{Aa} \\
&+ \tfrac{1}{4} Q_\mu{}^{IJ}(\Gamma^{IJ})_{AB}\chi_{Ba} + Q_{\mu ab}\chi_{Ab} + 2gA_\mu{}^{(r)}V_{Aa}{}^\alpha (D_\alpha(Q)\xi_{Bb}{}^{(r)})\chi_{Bb} \;. \quad (4.3)
\end{aligned}$$

As before, we use notations like $Q_\mu{}^{IJ} \equiv (\mathcal{D}_\mu \phi^\alpha) Q_\alpha{}^{IJ}$. The Killing vectors $\xi^{\alpha\,(r)}$ are in the directions of an arbitrary gauged group $H_0 \subset SO(8,n) \times SO(n)$ with $(r), (s), \cdots = (1), (2), \cdots, (d)$ for $d \equiv \dim H_0$. The relationship corresponding to (3.11) is now

$$V_{Aa}{}^\alpha D_\alpha(Q)\, \xi_{Bb}{}^{(r)} = +\tfrac{1}{4}(\Gamma^{IJ})_{AB} \delta_{ab} Q_\alpha{}^{IJ} \xi^{\alpha(r)} + \delta_{AB} Q_{\alpha ab} \xi^{\alpha(r)} \ . \tag{4.4}$$

As is easily seen, even though there is an additional factor group in $SO(8,n)/SO(n) \times SO(8)$, it exhibits a parallel structure between (4.1) and (3.6). Thus we can treat both of these cosets in a more unified fashion like that in [3], the details of which we skip in this paper due to space limitation.

Note that the case with $SO(8,n)/SO(n) \times SO(8)$ can not come from the $N=8$ supergravity [1][2] via any truncations. The reason is that in our theory $\chi_{A\alpha}$ belongs to the $\mathbf{8}_S$ of $SO(8)$, while $\phi^\alpha$ is equivalent to $\varphi_{Aa}$ in the $\mathbf{8}_S$ of $SO(8)$. In other words, we have the $(\varphi, \chi)$ in the $(\mathbf{8}_S, \mathbf{8}_S)$ of $SO(8)$, while those in [1][3] are either $(\mathbf{8}_V, \mathbf{8}_C)$ or $(\mathbf{8}_S, \mathbf{8}_C)$ because of the 'triality' of $SO(8)$, but not $(\mathbf{8}_S, \mathbf{8}_S)$ due to the chirality-flipping by the $\Gamma$-matrix $(\Gamma^I)_{A\dot A}$ [3]. In particular, for the above-mentioned reason, both fermions and bosons in $N=8$ in [1] can not be put into the same representations like ours. Notice also that this situation is different from the case of $F_{4(-20)}/SO(9)$, in which $N=9$ supergravity with $F_{4(-20)}/SO(9)$ has the same Majorana $\mathbf{16}$-representation for the whole $\sigma$-model multiplet, due to the oddness of 9 in $SO(9)$. (Cf. Table 1.) In other words, $N=9$ supergravity with $F_{4(-20)}/SO(9)$ can reproduce $N=1$ supergravity with the same coset by some truncations.

## 5. Globally $N=1$ Supersymmetric $\sigma$-Model on $E_{8(+8)}/SO(16)$

As we promised, we next clarify the realization of these $\sigma$-models only with *global* supersymmetry without supergravity. Even though this looks rather straightforward, after we have constructed the local case, the consequences seem non-trivial for two reasons. The first reason is that for general supergravity theories, there has been another wisdom about the couplings to supergravity restricting the algebraic structure of $\sigma$-model cosets. A typical example is $N=2$ supersymmetric $\sigma$-model in 4D. It has been well-known [4] that *global* $N=2$ supersymmetry in 4D requires the coset to be hyper-Kähler manifold, which is further restricted to quaternionic Kähler manifold, when coupled to $N=2$ supergravity. From this example, we expect that the same coset may not be consistent, if couplings to supergravity are turned off, when going to global supersymmetry. In 3D, however, we see that the coset structure of $E_{8(+8)}/SO(16)$ is 'independent' of the couplings to supergravity, *i.e.*, the same coset with exactly the same field representations can be supersymmetrized even with *global* $N=1$ supersymmetry. The second reason is related to the possible link with M-theory [5][7], or the question between global and local supersymmetrization of the maximal coset coming from 11D supergravity [8], as mentioned in the Introduction.

The corresponding lagrangian is easily constructed by truncating all the supergravity fields in section 2. However, one caveat is that when we study the $\chi^4$-quartic terms, there arises certain subtlety related to the Fierz rearrangements. Our lagrangian for $N = 1$ globally supersymmetric $\sigma$-model on $G/H = E_{8(+8)}/SO(16)$ or $F_{4(-20)}/SO(9)$ thus-obtained is

$$\begin{aligned}\mathcal{L}_{\text{global G/H}} &= + \tfrac{1}{4} g_{\alpha\beta} (\partial_\mu \phi^\alpha)(\partial^\mu \phi^\beta) - \tfrac{i}{2}(\overline{\chi}_A \gamma^\mu \mathcal{D}_\mu(Q) \chi_A) + \tfrac{1}{12} R_{ABCD} (\overline{\chi}_A \gamma_\mu \chi_B)(\overline{\chi}_C \gamma^\mu \chi_D) \\ &\quad - \tfrac{1}{4}(F_{\mu\nu}{}^{(r)})^2 - i(\overline{\lambda}^{(r)} \gamma^\mu D_\mu(A) \lambda^{(r)}) \\ &\quad + \tfrac{1}{2} c\, m\, \epsilon^{\mu\nu\rho} (F_{\mu\nu}{}^{(r)} A_\rho{}^{(r)} - \tfrac{1}{3} g f^{(r)(s)(t)} A_\mu{}^{(r)} A_\nu{}^{(s)} A_\rho{}^{(t)}) + 2 c\, m\, (\overline{\lambda}^{(r)} \lambda^{(r)})\ ,\end{aligned} \quad (5.1)$$

with global $N=1$ supersymmetry

$$\begin{aligned}\delta_Q \phi^\alpha &= V_A{}^\alpha (\overline{\epsilon} \chi_A)\ , &\delta_Q \chi_A &= +\tfrac{i}{2} \gamma^\mu \epsilon V_{\alpha A} \partial_\mu \phi^\alpha - \tfrac{1}{4}(\delta_Q \phi^\alpha) Q_\alpha{}^{IJ} (\Gamma^{IJ} \chi)_A\ , \\ \delta_Q A_\mu{}^{(r)} &= +i(\overline{\epsilon} \gamma_\mu \lambda^{(r)})\ , &\delta_Q \lambda^{(r)} &= -\tfrac{1}{4} \gamma^{\mu\nu} \epsilon F_{\mu\nu}{}^{(r)}\ .\end{aligned} \quad (5.2)$$

It is interesting to see the absence of $(\overline{\chi}\chi)^2$-term that was present in the local case (2.4). This can be traced back to the absence of the $\psi$-torsion in the variation of the $\chi$-field kinetic term. Needless to say, we do not have the cosmological constant, when supersymmetry is realized globally.

As for 'external' gaugings, there seems to be a problem related to the absence of the Noether term. In the local case (3.6), the $g\chi F$-terms out of the variations of the Noether term and $g(\chi\lambda)$-term cancelled each other. However, in the global case, the absence of the former requires the absence of the latter, which in turn implies that the $g\lambda\partial\phi$ terms in $\delta_Q \mathcal{L}$ have no counter-term, indicating the failure of the minimal coupling.

This result leads to the natural conjecture that a similar procedure can be performed to get globally $N = 1$ supersymmetric $\sigma$-models for other cosets in Table 1, as well as $SO(8,n)/SO(8) \times SO(n)$, whose confirmation we skip in this paper.

## 6. Concluding Remarks

In this paper, we have carried out an explicit construction of $N = 1$ supergravity in 3D coupled to a supersymmetric $\sigma$-model on the 'maximal' coset $E_{8(+8)}/SO(16)$ with both bosons and fermions in the same chiral **128**-representation of $SO(16)$, which is not covered as a special case in the unified formulation in [3]. We have also seen that such a supersymmetric $\sigma$-model is possible even without coupling to supergravity.

The results in our present paper elucidate several important new aspects of supergravity in 3D: First, we can actually couple supergravity to $\sigma$-model on huge cosets like $E_{8(+8)}/SO(16)$ with the $\sigma$-model multiplet in the $(\mathbf{128}, \mathbf{128})$ of $SO(16)$, which can not be obtained from $N = 16$ supergravity [1][3][2] by truncations. This is due to the difference of the field representations $(\mathbf{128}, \overline{\mathbf{128}})$ in the latter. Second, we can apply this method to other cosets like those in Table 1. As an explicit example, we have dealt with the metric notation compatible with $F_{4(-20)}/SO(9)$. Third, we can also add some cosmological constant and gravitino mass term, as in $N=16$ supergravity [2]. Fourth, we have seen that the 'internal' gauging is not possible for any subgroup in the same fashion as in [2]. However, since we can choose such a wide variety of cosets within $N=1$ supergravity as in Table 1, we do not regard this as a drawback. Fifth, we have seen that we can introduce an additional vector multiplet for 'external' gauging, with no restriction on allowed gauge groups for the subgroups $H_0 \subset H$ of $G/H$ for $E_{8(+8)}/SO(16)$ or $F_{4(-20)}/SO(9)$. We can also add a Chern-Simons term for the vector multiplet, leading to the 'generalized self-duality' field equation. Sixth, we have further seen that the $N=1$ globally supersymmetric $\sigma$-model on the coset $E_{8(+8)}/SO(16)$ is realized without supergravity, and the same seems also true for other cosets in Table 1, and $SO(8,n)/SO(8) \times SO(n)$.

Our result in this paper also constitutes a good counter-example against the conventional wisdom that since supergravity in 11D [8] with $128+128$ degrees of freedom naturally yields the maximal supergravity in dimensions $D \leq 10$, and that any other 'lower' $N$ supergravity within that dimension is most likely obtained by truncations from the maximal supergravity. Hence, any lower $N$ supergravity has necessarily 'fewer' degrees of freedom. Therefore, some degrees of freedom in $128+128$ in the maximal $N=16$ supergravity in 3D are supposed to be lost in the truncation to reach our $N=1$ supergravity. However, as we have seen, the total degrees of freedom of our $N=1$ supergravity are still $128+128$ in the irreducible representations of $SO(16)$, with the total number maintained to be the same as $N=16$ supergravity [1][3][2].

Note also that our $N=1$ supergravity with $E_{8(+8)}/SO(16)$ is *not* obtained by a truncation from the maximal $N=16$ supergravity in 3D [1][3][2]. The reason is that our $\sigma$-model fields $(\varphi_A, \chi_A)$ are both in the same chiral **128** of $SO(16)$, while those in [1][3][2] have $(\varphi_A, \chi_{\dot{A}})$ in the chiral **128** and anti-chiral $\overline{\mathbf{128}}$-representations. Even though our $N=1$ theory has obviously fewer supersymmetries compared with the maximal $N=16$ [1][2], our system has different representations, and yet the same $128+128$ physical degrees of freedom. The fact that our $N=1$ system does not come from $N=16$, but has the same degrees of freedom $128+128$ is peculiar to 3D, where supergravity has no physical degree of freedom. From these considerations, we may regard our $N=1$ supersymmetric $\sigma$-model on $E_{8(+8)}/SO(16)$ as the 'chiral' version of $E_{8(+8)}/SO(16)$, while that from $N=16$ [1][3][2] as a 'non-chiral' version. We can further regard this new feature as 'duality' or even 'triality' for all the cosets in Table 1.[5]

It has been known that any Riemannian manifold can be the consistent target space for locally $N=1$ supersymmetric $\sigma$-models in 3D [3]. Since the coset $E_{8(+8)}/SO(16)$ is also a Riemannian manifold, it is no wonder that $N=1$ supergravity can be coupled to $E_{8(+8)}/SO(16)$ itself. However, the important point is that our fermions $\chi$ are in the **128** of $SO(16)$ instead of the $\overline{\mathbf{128}}$ considered in the construction in [3]. In other words, our result has opened a new direction for possible $\sigma$-model representations in $N=1$ supergravity.

Our result makes sense also from the following viewpoint: Since there is no physical degree of freedom for the supergravity multiplets in 3D, 'extended supersymmetries' do not have such strong significance as in 4D or higher. Once we recognize this crucial point, it is not a far-fetched wishful-thinking to expect that our $N=1$ supergravity has an equally good chance of being directly related to M-theory [5][7] in 11D [8] as the maximal $N=16$ supergravity [1][3][2].

---
[5]This is with the exception of $F_{4(-20)}/SO(9)$ due to the same chirality shared by bosons and fermions.

It seems also true that our $N=1$ supergravity with $E_{8(+8)}/SO(16)$ has no higher-dimensional 'ancestor' theory among conventional ones, such as 11D supergravity [8]. This is inferred from the fact that our multiplet $(\varphi_A, \chi_A)$ forms the 'irreducible' $\mathbf{128}+\mathbf{128}$ of $SO(16)$ with the coset $E_{8(+8)}/SO(16)$, whose higher-dimensional origin is not clear. This is in a sharp contrast to the multiplet $(\varphi_A, \chi_{\dot{A}})$ in the maximal $N=16$ supergravity [1][2] with its direct origin in 11D supergravity [8]. Thus our $N=1$ $E_{8(+8)}/SO(16)$ supergravity seems to be disconnected from the conventional higher-dimensional theories, as well as from other higher $N$ within 3D [3]. Note also that the 'minimal' supergravity by simple dimensional reduction from $N=1$ supergravity in 4D is $N=2$ supergravity in 3D. In this sense, our $N=1$ supergravity in 3D is more 'chiral', like $N=1$ supergravity in 10D.

We have also emphasized the connection between the global and local supersymmetries, such as the same coset $E_{8(+8)}/SO(16)$ realized both with $N=16$ local and $N=1$ global supersymmetries. A close link between global and local supersymmetries for the same coset may be the first manifestation of their important relationship with M-theory [5][7], which is yet to be uncovered. The AdS/CFT correspondence between Type IIB supergravity and a globally $N=4$ supersymmetric Yang-Mills theory in 4D [9][10][11] is another suggestive example in this direction. The redundancy of local supersymmetry for supersymmetric $\sigma$-models suggests a more direct link between local supergravity with AdS(3) and global supersymmetric theories already within 3D before going down to 2D.

It is a peculiar nature in 3D that there is certain freedom for supersymmetric theories. This can be observed roughly in two different directions. The first one is, as we have seen, so much freedom of constructing $\sigma$-models within $N=1$ supergravity, where we can accommodate even the 'largest' coset $E_{8(+8)}/SO(16)$. The second direction is to go to higher values of $N$ up to $N\to\infty$ with no limit, in terms of Chern-Simons formulations [18], called $\aleph_0$ supergravity theories [19], in the absence of $\sigma$-models. Even though these two directions seem complementary to each other, they may well be the manifestation of different phases of a more fundamental theory such as M-theory [5][7][6]. Or, turning the table around, we can use 3D supergravity as a 'working ground' for the better understanding of M-theory itself [5][7][6].

We believe that our result in this paper has the potential to generate other supergravity models with $N=1$ supergravity in 3D that have not been explored in the past. We emphasize that the crucial ingredient in our work is the fact that in 3D the supergravity multiplets have no physical degree of freedom. This fact is in sharp contrast with what we know in higher dimensions.

We are grateful to P.K. Townsend for informing us about supersymmetric $\sigma$-models in 3D.